\documentclass[12pt,letter]{article}

\usepackage{graphicx, epsfig, color}
\usepackage{amssymb}
\def\eslt{\not\!\!{E_T}}
\def\delew{\Delta_{EW}}
\def\delhs{\Delta_{HS}}
\def\delbg{\Delta_{BG}}
\def\to{\rightarrow}

\def\bi{\begin{itemize}}
\def\ei{\end{itemize}}

\def\tb{\tilde b}

\def\tst{\tilde t}

\def\tg{\tilde g}

\def\tq{\tilde q}
\def\tw{\widetilde W}
\def\tz{\widetilde Z}
\def\be{\begin{equation}}  
\def\ee{\end{equation}}  
\def\bea{\begin{eqnarray}}  
\def\eea{\end{eqnarray}}  
\def\beas{\begin{eqnarray*}}  
\def\eeas{\end{eqnarray*}}  
\def\alt{\stackrel{<}{\sim}}
\def\agt{\stackrel{>}{\sim}}
\newcommand\prd[3]{{\it Phys.\ Rev.\ }{\bf D #1} (#2) #3}

\newcommand\prl[3]{{\it Phys.\ Rev.\ Lett.\ }{\bf #1} (#2) #3}
\newcommand\plb[3]{{\it Phys.\ Lett.\ }{\bf B #1} (#2) #3}
\newcommand\jhep[3]{{\it J. High Energy Phys.\ }{\bf #1} (#2) #3}

\newcommand\npb[3]{{\it Nucl.\ Phys.\ }{\bf B #1} (#2) #3}

\newcommand\ptp[3]{{\it Prog.\ Theor.\ Phys.\ }{\bf #1} (#2) #3}

\newcommand{\hepph}[1]{hep-ph/#1}

\newcommand\ppnp[3]{{\it Prog.\ Part.\ Nucl.\ Phys.}{\bf  #1} (#2) #3}

\begin{document}
\begin{titlepage}
\begin{flushright}
OU-HEP-130931
\end{flushright}

\vspace{0.5cm}
\begin{center}
{\Large \bf How conventional measures overestimate\\
electroweak fine-tuning in supersymmetric theory}
\\ 
\vspace{1.2cm} \renewcommand{\thefootnote}{\fnsymbol{footnote}}
{\large Howard Baer$^1$\footnote[1]{Email: baer@nhn.ou.edu },
Vernon Barger$^2$\footnote[2]{Email: barger@pheno.wisc.edu }
and Dan Mickelson$^1$\footnote[3]{Email: mickelso@nhn.ou.edu } 
}\\ 
\vspace{1.2cm} \renewcommand{\thefootnote}{\arabic{footnote}}
{\it 
$^1$Dept. of Physics and Astronomy,
University of Oklahoma, Norman, OK 73019, USA \\
}
{\it 
$^2$Dept. of Physics,
University of Wisconsin, Madison, WI 53706, USA \\
}

\end{center}

\vspace{0.5cm}
\begin{abstract}
\noindent 
The lack of evidence for superparticles at the CERN LHC, along with the
rather high value of the Higgs boson mass, has sharpened the perception
that what remains of supersymmetric model parameter space suffers a high
degree of electroweak fine-tuning (EWFT).  
We compare three different measures of fine-tuning in supersymmetric models.
 1. $\Delta_{HS}$ measures a subset of terms containing large
log contributions to $m_Z$ (and $m_h$) that are inevitable in models
defined at scales much higher than the electroweak scale. 
2. The traditional $\Delta_{BG}$ measures fractional variation in $m_Z$
against fractional variation of model parameters and allows for
{\it correlations} among high scale parameters which are not
included in $\delhs$.  
3. The model-independent $\Delta_{EW}$ measures how naturally a model can generate the measured value of 
$m_Z=91.2$~GeV (or $m_h$) in terms of {\em weak scale parameters} alone. 
We hypothesize an overarching Ultimate Theory (UTH) wherein the high scale 
soft terms are all correlated. 
The UTH might be contained within the more general effective SUSY theories which are 
popular in the literature.
In the case of $\Delta_{HS}$, EWFT can be grossly overestimated by neglecting additional 
non-independent terms which lead to large cancellations. In the case of $\Delta_{BG}$, 
EWFT can be overestimated by applying the measure to the effective theories instead of to the UTH. 
The measure $\delew$ allows for the possibility of parameter correlations which should be present 
in the UTH and, since it is model-independent, provides the same value of EWFT for the
effective theories as should occur for the UTH.
We find that the well-known mSUGRA/CMSSM model is fine-tuned under all three measures 
so that it is unlikely to contain the UTH. 
The non-universal Higgs model NUHM2
appears fine-tuned with $\Delta_{HS,BG} \gtrsim 10^3$. But since 
$\delew$ can be as small as 7 (corresponding to 14\% fine-tuning), it may contain the UTH for 
parameter ranges which allow for low true EWFT.

\vspace*{0.8cm}

\end{abstract}

\end{titlepage}

\section{Introduction} 
\label{sec:intro}

The recent discovery of a Standard Model (SM)-like Higgs boson with mass 
$m_h=125.5\pm 0.5$~GeV\cite{atlas_h,cms_h} at the LHC
seemingly provides credence to the simplest SUSY
models of particle physics\cite{wss,books} which had predicted $m_h\alt 135$~GeV\cite{mhiggs}.
On the other hand, no sign of supersymmetric matter has yet emerged at the LHC, leading to mass
limits $m_{\tg}\gtrsim 1.5$~TeV (for $m_{\tg}\simeq m_{\tq}$) and
$m_{\tg}\gtrsim 1$~TeV (for $m_{\tg}\ll m_{\tq}$)\cite{atlas_susy,cms_susy}. 
These limits, obtained within the context of popular models such as mSUGRA/CMSSM\cite{msugra} 
or simplified models,  are qualitatively also valid in many other frameworks 
as long as we understand that the squark mass limit refers to first generation squarks. 
The squark and gluino mass limits have caused considerable concern since for many years 
the storyline has been promoted that in order to maintain naturalness in SUSY models, 
sparticles ought to be well below the 
TeV scale~\cite{ellis,bg,kane,ac1,dg,ccn,ellis2,king,casas,ross,shafi,perel,antusch,hardy,sug19,Fichet:2012sn,Kowalska:2013ica,han,Dudas:2013pja,Arvanitaki:2013yja}. 
Indeed, the absence of any hint of deviations from the SM in the LHC8 data have
led some to question whether SUSY could be the solution to the naturalness
problem of the SM.

The fine-tuning situation in the Minimal Supersymetric Standard Model (MSSM) is further 
exacerbated by the uncomfortably large value of the newly discovered Higgs particle:
its value $m_h\simeq 125$ GeV lies well beyond its tree-level upper bound $m_h \le m_Z$. 
Radiative corrections can accommodate $m_h\simeq 125$~GeV but only at the expense of 
either 1. having top squark masses beyond the TeV scale along with large mixing\cite{h125}, or else
2. enlarging the MSSM to contain additional contributions to $m_h$\cite{hall,bae,e6}. 
The first of these possibilities again seems in violation of naturalness limits which 
according to many studies require $m_{\tst_{1,2}},m_{\tb_1}\lesssim 500$~GeV\cite{kn,ah,sundrum,ns}.

Thus, the question arises: 
are SUSY models now unnatural, and if so, how unnatural are they? 
Or, do there exist portions of parameter space where SUSY remains natural? 
If so, a credible goal of collider\cite{lhcltr,gurrola} and dark matter\cite{bbm} search experiments is
to conduct a thorough search for natural SUSY. 

In this paper, we compare and contrast three different measures of SUSY naturalness:
 1. $\Delta_{HS}$ measures a subset of terms containing large
log contributions to $m_Z$ (and $m_h$) that are inevitable in models
defined at scales much higher than the electroweak scale. 
2. The traditional $\Delta_{BG}$ measures fractional variation in $m_Z$
against fractional variation of model parameters and allows for
{\it correlations} among high scale parameters which are not included in $\delhs$.  
3. The model-independent $\Delta_{EW}$ measures how naturally a model can generate the measured value of 
$m_Z=91.2$~GeV (or $m_h$) in terms of {\em weak scale parameters} alone. 
Low values of $\Delta_i$ ($i=HS,\ BG$ or $EW$) mean low fine-tuning, {\it e.g.} $\Delta_i =100$ corresponds to 
$\Delta_i^{-1}=1\%$ EWFT.

For illustrative purposes, we apply these measures to two popular high scale SUSY
models: the paradigm mSUGRA/CMSSM model\cite{msugra} based on the parameter set 
\be
m_0,\ m_{1/2},\ A_0,\ \tan\beta ,\ sign(\mu ) 
\ee 
and the more general two-extra-parameter non-universal Higgs
model NUHM2\cite{nuhm2} defined by the parameter set 
\be
m_0,\ m_{1/2},\ A_0,\ \tan\beta ,\ \mu ,\ m_{A} 
\ee 
(where we have traded the GUT scale soft SUSY
breaking Higgs mass parameters $m_{H_u}^2$ and $m_{H_d}^2$ for the
weak scale parameters $\mu$ and $m_A$ for convenience).
We find the measures ordered according to
\be
\Delta_{EW}<\Delta_{BG}\alt \Delta_{HS} .
\ee
We argue that the semi-model-independent $\Delta_{HS}$ omits non-independent terms from its measure 
which lead to large cancellations giving rise to an overestimate of EWFT.
The measure $\Delta_{BG}$ properly combines these correlated terms so as to avoid the pitfall contained within $\Delta_{HS}$. 
To interpret $\Delta_{BG}$ properly, we hypothesize an overarching Ultimate Theory 
whose low energy limit for $Q<\Lambda =m_{GUT}$ is the MSSM wherein the high scale 
soft terms are all {\it correlated} (hereafter referred to as the UTH). 
The UTH might be contained within the more general effective SUSY theories which are popular in the literature.
Examples include the mSUGRA/CMSSM model and the NUHM2 model.
In the case of $\Delta_{BG}$, EWFT can be overestimated by applying the measure to the 
effective theories instead of to the UTH. 
The measure $\delew$ allows for the possibility of parameter correlations which should be present 
in the UTH and, since it is model-independent, provides the same value of $\delew$ for the
effective theories as should occur for the UTH.
We find that the well-known mSUGRA/CMSSM model is fine-tuned under all three measures 
so that it cannot contain the UTH. The non-universal Higgs model NUHM2
appears fine-tuned with $\Delta_{HS,BG} \gtrsim 10^3$. But since 
$\delew$ can be as small as 7 (corresponding to 14\% fine-tuning), it may contain the UTH for
the range of parameter choices which allow for low true EWFT. 

The low $\Delta_{EW}$ models 
are characterized by a superpotential $\mu$ term with $|\mu |\sim m_Z\sim 100-300$ GeV. 
This leads to a prediction of light higgsino states $\tw_1^\pm$, $\tz_2$ and $\tz_1$ with mass
$\sim 100-300$ GeV which, due to their compressed spectra, may easily elude LHC searches, but which
should be accessible to an $e^+e^-$ collider with $\sqrt{s}\agt 2|\mu |\sim 500-600$ GeV.

In Sec.~\ref{sec:measures}, we define and review the measures 
$\Delta_{HS}$, $\Delta_{BG}$ and $\Delta_{EW}$ which were mentioned above. 
In Sec.~\ref{sec:sugra}, we evaluate the three measures
as a function of parameters in the mSUGRA/CMMSM model. 
We repeat our evaluation for the NUHM2 model in Sec.~\ref{sec:nuhm}. 
In Sec. \ref{sec:meta} we interpret our results in terms of an overarching UTH.
In Sec.~\ref{sec:concl} we present our general conclusion which is that the conventional measures of 
naturalness $\Delta_{HS}$ and $\Delta_{BG}$ lead to large overestimates of EWFT in supersymmetric theory. 
Parameter choices exist within the
NUHM2 model (and of course other more general models) which lead to low $\Delta_{EW}$ and about one part in ten
EWFT. Such parameter choices should be a guide to model builders seeking to find the correct UTH which
predicts their values in terms of few or even no adjustable parameters.\footnote{An example along these lines 
is provided in Ref.~\cite{cheng}.}

\section{Three fine-tuning measures} 
\label{sec:measures}

\subsection{$\Delta_{HS}$}

\subsubsection{Standard Model}

In the SM, with a Higgs potential given by $V=-\mu^2\phi^\dagger\phi+\lambda (\phi^\dagger\phi )^2$ where 
$\phi$ is the Higgs doublet, one may calculate the physical mass of the Higgs boson $m_h$ as
\be
m_h^2  =m_h^2|_{tree}+\delta m_h^2|_{rad}
\label{eq:mhSM}
\ee 
where $m_h^2|_{tree}=\sqrt{2}\mu^2$ and $\delta m_h^2|_{rad}=\frac{c}{16\pi^2}\Lambda^2$, 
and where $\Lambda$ represents the cutoff of quadratically divergent loop diagrams which provides 
an upper limit to which the SM is considered a valid effective field theory. 
The coefficient $c$ depends on the various SM couplings and here will be taken
$c\sim 1$ ({\it e.g.} the top quark loop gives $c=-6f_t^2$ where $f_t$ is the top quark Yukawa coupling).
Since $m_h^2|_{tree}$ and $\delta m_h^2|_{rad}$ are {\it independent}, we would expect naturally that  
$m_h^2 \sim m_h^2|_{tree} > \delta m_h^2|_{rad}$ since otherwise if $ \delta m_h^2|_{rad}\gg m_h^2$ then
$m_h^2|_{tree}$ will have to be fine-tuned to a high degree to obtain $m_h$ of just $\sim 125$ GeV.
We may define a fine-tuning measure
\be
\Delta_{SM}\equiv \delta m_h^2|_{rad}/(m_h^2/2)
\ee
which compares the radiative correction to the physical Higgs boson mass.
Requiring $\Delta_{SM} \alt 1$ then requires $\Lambda \sim 1$ TeV, {\it i.e.} the SM should only be valid
up to at most the TeV scale.

\subsubsection{MSSM}

Analogous reasoning has been applied to supersymmetric models\cite{kn}. 
In the MSSM, then
\be
m_h^2 \simeq \mu^2 +m_{H_u}^2 +\delta m_{H_u}^2|_{rad} 
\ee 
where now $\mu$ is the superpotential Higgs/higgsino mass term and $m_{H_u}^2$ is the up-type
soft SUSY breaking Higgs mass evaluated at $m_{SUSY}\sim 1$ TeV.
In gravity-mediation, then $m_{H_u}$ is expected $\sim m_{3/2}\sim 1$ TeV.
The largest contributions to $\delta m_{H_u}^2|_{rad}$ contain divergent logarithms; these can be found
by integrating the renormalization group equation\cite{bbo} for $m_{H_u}^2$:
\be
\frac{dm_{H_u}^2}{dt}=\frac{1}{8\pi^2}\left(-\frac{3}{5}g_1^2M_1^2-3g_2^2M_2^2+\frac{3}{10}g_1^2 S+3f_t^2 X_t\right)
\ee
where $t=\ln (Q^2/Q_0^2)$, $S=m_{H_u}^2-m_{H_d}^2+Tr\left[{\bf m}_Q^2-{\bf m}_L^2-2{\bf m}_U^2+{\bf m}_D^2+{\bf m}_E^2\right]$
and where $X_t=m_{Q_3}^2+m_{U_3}^2+m_{H_u}^2+A_t^2$.
By neglecting gauge terms and $S$ ($S=0$ in models with scalar soft term universality), 
and also neglecting the $m_{H_u}^2$ contribution to $X_t$ and the fact that $f_t$ and the soft terms
evolve under $Q^2$ variation, 
then this expression may be readily integrated from $m_{SUSY}$ to the cutoff $\Lambda$ to obtain 
\be
\delta m_{H_u}^2|_{rad}\sim -\frac{3f_t^2}{8\pi^2}(m_{Q_3}^2+m_{U_3}^2+A_t^2)\ln\left(\Lambda^2/m_{SUSY}^2 \right) .
\label{eq:DBoE}
\ee
Inspired by gauge coupling unification, $\Lambda$ may be taken as high as $m_{GUT}\simeq 2\times 10^{16}$ GeV 
or even the reduced Planck mass $m_P\simeq 2.4\times 10^{18}$ GeV.
Also, we take $m_{SUSY}^2 \simeq m_{\tst_1}m_{\tst_2}$. 
One may again create a fine-tuning measure $\Delta \equiv \delta m_{H_u}^2/(m_h^2/2)$. 

Two related dangers are contained within this approach, which are different from the case of the SM. 
\begin{itemize} 
\item The first is that $m_{H_u}^2$ and $\delta m_{H_u}^2|_{rad}$ are not independent:
the value of $m_{H_u}^2$ feeds directly into evaluation of $\delta m_{H_u}^2|_{rad}$ via the $X_t$ term.
It also feeds indirectly into $\delta m_{H_u}^2|_{rad}$ by contributing to the evolution of the
$m_{Q_3}^2$ and $m_{U_3}^2$ terms. In fact, the larger the value of $m_{H_u}^2(\Lambda )$, then the
larger is the cancelling correction $\delta m_{H_u}^2|_{rad}$. We return to this issue later.
\item The second is that whereas $SU(2)_L\times U(1)_Y$ gauge symmetry can be broken at tree
level in the SM, in the SUSY case $m_{H_u}^2\sim m_{3/2}>0$ and EW symmetry is not even broken until
one includes radiative corrections. For high scale SUSY models, 
EW symmetry is broken radiatively by $m_{H_u}^2$ being driven to 
large negative values.
This suggests a re-grouping of terms\cite{ltr,rns}:
\be
m_h^2|_{phys}=\mu^2+\left(m_{H_u}^2(\Lambda )+\delta m_{H_u}^2 \right)
\label{eq:mh}
\ee
where instead both $\mu^2$ and $(m_{H_u}^2+\delta m_{H_u}^2)$ should be comparable to $m_h^2|_{phys}$.
\end{itemize}

Nonetheless, using the measure $\Delta$, Eq.~\ref{eq:DBoE} may be re-arranged to provide a
bound on third generation squarks\cite{kn,ah,ns}:
\be \sqrt{m_{\tst_1}^2+m_{\tst_2}^2} \alt 600 \ {\rm GeV}
\frac{\sin\beta}{\sqrt{1+R_t^2}}\left(\frac{\log\frac{\Lambda}{{\rm
TeV}}}{3}\right)^{-1/2}\left(\frac{\Delta}{5}\right)^{1/2}\;,
\label{eq:papucci}
\ee 
where $R_t= A_t/\sqrt{m_{\tst_1}^2+m_{\tst_2}^2}$.  Taking $\Delta =10$
({\it i.e.} $\Delta^{-1}=10\%$ EWFT) and $\Lambda$ as low as 20~TeV corresponds to
\bi
\item $m_{\tst_i},\ m_{\tb_1}\alt 600\ {\rm GeV}$,
\item $m_{\tg}\alt 1.5-2\ {\rm TeV}$.
\ei
The last of these conditions arises because the squark radiative corrections
$\delta m_{\tst_i}^2\sim (2g_s^2/3\pi^2)m_{\tg}^2 \times \log\Lambda$.  Setting
the $\log$ to unity and requiring $\delta m_{\tst_i}^2<m_{\tst_i}^2$
then implies $m_{\tg}\alt 3m_{\tst_i}$, or $m_{\tg}\alt 1.5-2$~GeV for
$\Delta\alt 10$.  
Taking $\Lambda$ as high as $m_{GUT}$ leads to even tighter constraints:
$m_{\tst_{1,2}},m_{\tb_1}\alt 200$ GeV and $m_{\tg}\alt 600$ GeV, 
almost certainly in violation of LHC sparticle search constraints. 
Since (degenerate) first/second generation
squarks and sleptons enter the Higgs potential only at the two loop
level, these can be much heavier: beyond LHC reach and also possibly
heavy enough to provide a (partial) decoupling solution to the SUSY flavor and $CP$
problems\cite{dine}.

To bring the fine-tuning measure $\Delta$ into closer accord with the measures described below, we 
write it in terms of $m_Z^2/2$ instead of in terms of $m_h^2/2$. 
The minimization condition for the Higgs potential  $V_{\rm tree} + \Delta V$ in the MSSM reads
\be 
\frac{m_Z^2}{2} = \frac{m_{H_d}^2 + \Sigma_d^d -
(m_{H_u}^2+\Sigma_u^u)\tan^2\beta}{\tan^2\beta -1} -\mu^2 \;,
\label{eq:loopmin}
\ee 
where $\Sigma_u^u$ and $\Sigma_d^d$ are radiative corrections that
arise from the derivatives of $\Delta V$ evaluated at the minimum.
The radiative corrections $\Sigma_u^u$ and $\Sigma_d^d$ include contributions from
various particles and sparticles with sizeable Yukawa and/or gauge
couplings to the Higgs sector.
Expressions for the $\Sigma_u^u$ and $\Sigma_d^d$ are given in the Appendix of Ref. \cite{rns}.
We may include explicit dependence on the high scale $\Lambda$ at which the SUSY theory may be defined, 
by writing the {\it weak scale} parameters $m_{H_{u,d}}^2$  as 
\be
m_{H_{u,d}}^2= m_{H_{u,d}}^2(\Lambda) +\delta m_{H_{u,d}}^2; \ \ \ \
\mu^2=\mu^2(\Lambda)+\delta\mu^2\;,
\ee 
where
$m_{H_{u,d}}^2(\Lambda)$ and $\mu^2(\Lambda)$ are the corresponding
parameters renormalized at the high scale $\Lambda$. 
The $\delta m_{H_{u,d}}^2$ terms contain the $\log\Lambda$ dependence 
emphasized in constructs of natural SUSY models\cite{kn,ah,ns}. 
Thus, one obtains
\be 
\frac{m_Z^2}{2} = \frac{(m_{H_d}^2(\Lambda)+ \delta m_{H_d}^2 +
\Sigma_d^d)-(m_{H_u}^2(\Lambda)+\delta m_{H_u}^2+\Sigma_u^u)\tan^2\beta}{\tan^2\beta -1} 
-(\mu^2(\Lambda)+\delta\mu^2)\;.
\label{eq:mZs_hs}
\ee 
We can now define a fine-tuning measure that encodes the
information about the high scale origin of the parameters by requiring
that each of the terms on the right-hand-side of Eq.~(\ref{eq:mZs_hs}) 
(normalized to $m_Z^2/2$) be smaller than a value $\Delta_{HS}$. 
The high scale fine-tuning measure $\Delta_{HS}$ is thus defined to be
\be 
\Delta_{HS}\equiv max_i |B_i |/(m_Z^2/2)\;, 
\label{eq:hsft} 
\ee 
with $B_{H_d}\equiv m_{H_d}^2(\Lambda)/(\tan^2\beta -1)$ etc.
In models such as mSUGRA,  whose domain of validity extends to very high scales,
because of the large logarithms one would expect that the $B_{\delta H_u}$ contributions to 
$\Delta_{HS}$ would be the dominant term.

An advantage of $\Delta_{HS}$ over $\Delta$ is that the dominant term $B_{\delta H_u}$
is extracted now from the RGE solution and thus includes large logs arising from gauge terms 
as well as the effect of running parameters which are not contained in Eq. \ref{eq:DBoE}. 
However, it still maintains the split amongst the {\it dependent} terms $m_{H_u}^2(\Lambda )$ and
$\delta m_{H_u}^2$.\footnote{The possibility of models with low $\Delta_{HS}$ is explored 
within the context of GUT models with non-universal gaugino masses\cite{shafi} and
the 19-parameter SUGRA model\cite{sug19}. Low $\Delta_{HS}$ requires small $m_{H_u}(\Lambda = m_{GUT})$
and then minimal evolution of $m_{H_u}^2$ between $m_{GUT}$ and $m_{SUSY}$. The low $\Delta_{HS}$
models tend to have sub-TeV top-squarks which lead typically to large deviations in $BF(b\to s\gamma )$.}

\subsection{$\Delta_{BG}$}

The fine-tuning measure $\Delta_{BG}$ can be regarded as the traditional measure, in use now for over
25 years\cite{ellis,bg,dg}.
We start again with the scalar potential minimization condition (this time at tree level)
\be 
\frac{m_Z^2}{2} = \frac{m_{H_d}^2 - m_{H_u}^2 \tan^2\beta}{\tan^2\beta -1} -\mu^2\simeq -m_{H_u}^2-\mu^2
\label{eq:mZs}
\ee 
where the latter partial equality obtains for moderate-to-large $\tan\beta$ values.
The traditional measure is then defined as
\be
\Delta_{BG}\equiv max_i\left[ c_i\right]\ \ {\rm where}\ \ c_i=\left|\frac{\partial\ln m_Z^2}{\partial\ln a_i}\right|
=\left|\frac{a_i}{m_Z^2}\frac{\partial m_Z^2}{\partial a_i}\right|
\label{eq:DBG}
\ee
where the $a_i$ constitute the fundamental parameters of the model.
Thus, $\Delta_{BG}$ measures the fractional change in $m_Z^2$ due to fractional variation in 
high scale parameters $a_i$. 
The $c_i$ are known as {\it sensitivity co-efficients}\cite{feng}.

An advantage of $\Delta_{BG}$ over $\Delta_{HS}$ or $\Delta$ is that it maintains the correlation
between $m_{H_u}^2(\Lambda )$ and $\delta m_{H_u}^2$ by replacing 
$m_{H_u}^2 (m_{SUSY})= m_{H_u}^2(\Lambda )+\delta m_{H_u}^2$ by its expression in terms of high scale parameters.
To evaluate $\Delta_{BG}$, one needs to know the explicit dependence of $m_{H_u}^2$ and $\mu^2$ on the
fundamental parameters. Expressions can be gained by semi-analytic solutions to the
one-loop renormalization group equations (RGEs), as found for instance in Ref's \cite{munoz}.
In the case where $\tan\beta =10$, it is found in Ref's \cite{abe,martin,feng} that
\bea
-2\mu^2(m_{SUSY}) &=& -2.18\mu^2 \\
-2m_{H_u}^2(m_{SUSY})&=& 3.84 M_3^2+0.32M_3M_2+0.047 M_1M_3-0.42 M_2^2 \nonumber \\
& & +0.011 M_2M_1-0.012M_1^2-0.65 M_3A_t-0.15 M_2A_t\nonumber \\
& &-0.025M_1 A_t+0.22A_t^2+0.004 M_3A_b\nonumber \\
& &-1.27 m_{H_u}^2 -0.053 m_{H_d}^2\nonumber \\
& &+0.73 m_{Q_3}^2+0.57 m_{U_3}^2+0.049 m_{D_3}^2-0.052 m_{L_3}^2+0.053 m_{E_3}^2\nonumber \\
& &+0.051 m_{Q_2}^2-0.11 m_{U_2}^2+0.051 m_{D_2}^2-0.052 m_{L_2}^2+0.053 m_{E_2}^2\nonumber \\
& &+0.051 m_{Q_1}^2-0.11 m_{U_1}^2+0.051 m_{D_1}^2-0.052 m_{L_1}^2+0.053 m_{E_1}^2 ,
\label{eq:mHu}
\eea
where the parameters on the right-hand-side are evaluated at the GUT scale.
For different values of $\tan\beta$, then somewhat different relations are obtained.
At this point, the derivatives in Eq. \ref{eq:DBG} can be explicitly evaluated
so that $\Delta_{BG}$ can be easily computed.

\subsubsection{The importance of high-scale correlations}

An important difference between $\Delta_{HS}$ and $\Delta_{BG}$ is that the latter combines
the dependent terms $m_{H_u}^2(\Lambda )$ and $\delta m_{H_u}^2$ which were separated in $\Delta_{HS}$. 
Including these allows for {\it cancellations} between various terms which occur 
if certain correlations between HS parameters arise in the model under consideration. 
For instance, in lines 6 and 7 of Eq. \ref{eq:mHu}, if we impose 
\be
m_{Q_{1,2}}=m_{U_{1,2}}= m_{D_{1,2}}=m_{L_{1,2}}=m_{E_{1,2}} \equiv m_{16}(1,2)
\ee
as might be expected in an $SO(10)$ GUT theory, then each line collapses to
$\sim 0.007 m_{16}^2(1,2)$: the various terms now conspire via cancellations to yield much 
less fine-tuning than otherwise might be expected. 

More importantly, if 
\be
m_{H_u}^2=m_{H_d}^2=m_{16}^2(3)\equiv m_0^2
\ee
as is imposed in models with scalar mass universality, then lines 4 and 5 of Eq. \ref{eq:mHu} 
conspire to yield a 
term $\sim -0.017 m_0^2$, which again yields far less fine-tuning in the third generation sector 
than one might otherwise expect due to cancellations of terms, many of which  
contain the large logs which are measured by $\Delta_{HS}$.
This latter case is usually refered to as ``focus point SUSY''\cite{ccn,fp}: it provides a concrete example 
that in the case of very heavy top squarks, the fine-tuning which follows from $\Delta_{HS}$ may be a
large over-estimate.\footnote{Note that if $m_{H_u}^2(\Lambda )$ is subtracted out of Eq. \ref{eq:mHu}
(as is done in Eq. \ref{eq:DBoE}), then the nearly complete cancellation of 
Higgs and third generation soft terms will no longer occur.}
Further cancellations amongst terms in Eq. \ref{eq:mHu} can occur when the $A_t$ parameters obey
certain relations to $m_{1/2}$.
Thus, the allowance for cancellations in the log terms of $\Delta_{BG}$ gives rise
to the expectation that
\be
\Delta_{BG}\alt \Delta_{HS} .
\label{eq:inequal2}
\ee

\subsubsection{Model dependence of $\delbg$}

At this point it is important to note that while Eq. \ref{eq:mHu} provides a good example of how
large log and other cancellations can occur due to HS parameter correlations, it is not at all 
clear that usage of \ref{eq:mHu} is the correct way to proceed.
There is often dispute in the literature as to which parameters should be included in the set $a_i$ which
enters into the evaluation of $\delbg$. Surely the high scale
soft SUSY breaking parameters would be included, but should also {\it e.g.} the top quark Yukawa
coupling $f_t$, or other Yukawa or even gauge couplings be included?\footnote{
Also, different papers will use varying powers of parameters as fundamental
inputs. For instance, in mSUGRA, does one use $m_0$ or $m_0^2$? 
These differences lead to just factors of 2 in the evaluation of $\Delta_{BG}$.} 
Furthermore, shall one use the Lagrangian trilinear
soft SUSY breaking parameter $a_t$ as occurs in 
${\cal L}\ni a_t\epsilon_{ab}\tilde{Q}_3^a H_u^b\tilde{u}_{R3}^\dagger$
or the more common $A_t$ where $a_t=f_t A_t$? Different prescriptions as to what one includes in the  
``fundamental parameters'' $a_i$ will lead to different expressions for $m_Z^2$ in terms of the $a_i$.

A further concern with $\Delta_{BG}$ is that different models with exactly the same weak scale spectra
can give rise to wildly different values of $\Delta_{BG}$. 
We will see that in the HB/FP region\cite{fp} of the
mSUGRA model, $\Delta_{BG}$ can be greatly reduced due to $m_{H_u}=m_{H_d}\equiv m_0$ at the GUT scale. 
Yet, using the exact same input parameters within the NUHM2 model 
(or any other model with greater parameter freedom which contains mSUGRA as a subset), 
then the value of $\Delta_{BG}$ will be quite a bit larger. 
An example is given in Table \ref{tab:coeff} which lists the various sensitivity co-efficients of the 
$\Delta_{BG}$ measure for mSUGRA and for NUHM2, but where the mSUGRA output values of $\mu$ and $m_A$
are used as inputs to NUHM2. In this case, the two models have exactly the same weak scale spectra.
But due to the greater correlations amongst HS parameters present in mSUGRA, the value of $\Delta_{BG}$ has
dropped by an order of magnitude compared to NUHM2.
\begin{table}
\begin{center}
\begin{tabular}{|l|r|r|r|r|r|r|r|}
\hline
model & $c_{m_0}$  &  $c_{m_{1/2}}$ & $c_{A_0}$ & $c_\mu$ & $c_{H_u}$ &
$c_{H_d}$ & $\Delta_{BG}$ \\
\hline
\hline
mSUGRA    & 156 &  762 & 1540 & -25.1 & $--$ & $--$ & 1540 \\
NUHM2     & 16041 &  762 & 1540 & -25.1 & -15208 & -643.6 & 16041 \\
\hline
\end{tabular}
\caption{Sensitivity coefficients and $\Delta_{BG}$ for mSUGRA 
and NUHM2 model 
with $m_0=9993.4$ GeV, $m_{1/2}=691.7$ GeV, $A_0=-4788.6$ GeV and
$\tan\beta =10$. The mSUGRA output values of $\mu =309.7$ GeV and 
$m_A=9859.9$ GeV serve as NUHM2 inputs so that the two models have exactly the same weak scale spectra.
\label{tab:coeff}}
\end{center}
\end{table}

Here, it must be remembered that models like mSUGRA or NUHM2 {\it etc.} are to be regarded as {\it effective theories} valid up to
$\Lambda=M_{GUT}$, and where the parameters parametrize our ignorance of high scale physics such as the mechanism for SUSY breaking.
It is usually regarded that such SUSY GUT models are the low energy effective field theories of some more encompassing theory
(ultimate theory, perhaps string theory) where further parameter correlations are to be expected, or perhaps there are no free parameters.
In such a case, the effective theory may look fine-tuned while the high scale correlations present in the 
UTH lead to little or no fine-tuning.

The fundamental lesson here is that: examples exist where correlations amongst model parameters 
which are present in more restrictive theories, but not in the effective theory within which they are contained, 
lead to cancellations in contributions to EWFT. 
In such cases, one may gain a false impression as to the amount of EWFT needed in a theory.
Is one then to give up on EWFT as a guide to a supersymmetric theory? 

\subsection{$\Delta_{EW}$}

A less ambitious, more conservative and model-independent, fine-tuning measure has been advocated in 
Ref's \cite{ltr,sugra,rns}.\footnote{The importance of low $|\mu |\sim m_Z$ was emphasized in
Ref.~\cite{ccn}. 
Ref.~\cite{fp} also remarks that there be no large cancellation between $m_{H_u}^2$ and $\mu^2$. 
Ref's~\cite{golden,perel,nevzrov,guido} essentially adopt weak scale fine-tuning. 
Ref.~\cite{ltr} creates $\Delta_{EW}$ including radiative corrections and notes 
that large $A_t$ suppresses radiative corrections while lifting 
the value of $m_h$.}
Starting again with the scalar potential minimization condition, this time including 
radiative corrections, we have
\be 
\frac{m_Z^2}{2} = \frac{m_{H_d}^2+\Sigma_d^d - (m_{H_u}^2+\Sigma_u^u) \tan^2\beta}{\tan^2\beta -1} -\mu^2 .
\label{eq:mZsSig}
\ee 
Noting that all entries in Eq.~\ref{eq:mZsSig} are defined at the weak scale, 
the {\it  electroweak fine-tuning measure} 
\be 
\Delta_{EW} \equiv max_i \left|C_i\right|/(m_Z^2/2)\;, 
\ee 
may be constructed, where $C_{H_d}=m_{H_d}^2/(\tan^2\beta -1)$, $C_{H_u}=-m_{H_u}^2\tan^2\beta /(\tan^2\beta -1)$ and $C_\mu =-\mu^2$. 
Also, $C_{\Sigma_u^u(k)} =-\Sigma_u^u(k)\tan^2\beta /(\tan^2\beta -1)$ and $C_{\Sigma_d^d(k)}=\Sigma_d^d(k)/(\tan^2\beta -1)$, 
where $k$ labels the various loop contributions included in Eq. \ref{eq:mZsSig}.

Constructed in this way, it is clear that
\be
\lim_{\Lambda\to m_{SUSY}} \Delta_{HS} =\Delta_{EW} . 
\ee
It can also be checked that
\be
\lim_{\Lambda\to m_{SUSY}} \Delta_{BG} \sim \Delta_{EW} , 
\ee
since the most important terms in Eq. \ref{eq:mZsSig} appear linearly in $m_{H_u}^2$ and $\mu^2$.
Thus, we expect that
\be
\Delta_{EW}<\Delta_{BG}\alt \Delta_{HS}
\label{eq:inequal}
\ee
for any particular point in a given model parameter space. 

The measure $\Delta_{EW}$ is created from weak scale SUSY parameters and so contains no information 
about any possible high scale origin: hence its model-independence. 
Since it evaluates the fine-tuning which remains upon taking the limit 
$\Lambda\to m_{SUSY}$, it makes an allowance for cancellations of large logs which may enter into 
$m_{H_u}^2(m_{SUSY})$. 
In this sense, $\Delta_{EW}$ captures the {\it minimal} amount of EWFT required of any SUSY
model, including those defined at some high scale $\Lambda\gg m_{SUSY}$.
$\delew$ can be thought of as providing a {\it lower bound} on electroweak fine-tuning\cite{snow_wp1}. 
Any model with a large value of $\Delta_{EW}$ is always fine-tuned.
However, if $\Delta_{EW}$ is low, it need not mean the model is not fine-tuned: 
rather, it allows for the possibility that some model might exist 
with low fine-tuning which might be hidden by the naive application of either $\Delta_{HS}$ or $\Delta_{BG}$.
As such, low $\Delta_{EW}$ is a {\it necessary, albeit not sufficient}, measure of electroweak fine-tuning.

The quantity  $\Delta_{EW}$ measures the largest {\it weak scale} contribution to the $Z$ mass. 
Model parameter choices which lead to low values of $\Delta_{EW}$ are those which would naturally
generate a value of $m_Z\sim 91.2$ GeV. 
In order to achieve low $\delew$,  it is necessary that $-m_{H_u}^2$, $\mu^2$ and $-\Sigma_u^u$ 
all be nearby to $m_Z^2/2$ to within a factor of a few\cite{ltr,rns}:
The low $\Delta_{EW}$ models are typified by the presence of light higgsinos $\tw_1^\pm$, $\tz_{1,2}$
with mass $\sim |\mu |\sim 100-300$ GeV.

\subsubsection{The utility of $\delew$}

We have emphasized that $\delew$ is a measure of the {\it minimal fine-tuning} 
that is present in a given weak scale SUSY spectrum. 
While a model with a small value of $\delew$ is not necessarily free of fine-tuning, 
any model with a large value of $\delew$ is always fine-tuned. 

The utility of $\delew$ arises from the fact that it is determined by just the weak scale spectrum\cite{rns}: 
{\it i.e.} different high scale theories that lead to the same sparticle spectrum will yield
the same value of $\delew$, even though these may have vastly different values of $\delhs$ or $\delbg$. 
A small value of $\delew$ in some region of parameter space of a SUSY effective theory offers the possibility
that there may exist an overarching UTH with essentially the same spectrum 
but whose parameter correlations lead to small values of $\delbg$. 
This UTH would then be the underlying theory with low true EWFT. 
Since the broad features of the phenomenology are determined by the spectrum, 
we expect that the phenomenological consequences of the (unknown) UTH
will be the same as for the more general effective theory which includes the UTH as a special case.

\section{The $\Delta_i$ in the mSUGRA/CMSSM model}
\label{sec:sugra}

To calculate superparticle mass spectra in SUSY models, we employ the
Isajet 7.83~\cite{isajet} SUSY spectrum generator Isasugra\cite{isasugra}. 
We begin with a scan over mSUGRA/CMSSM parameter space for a fixed value of
$\tan\beta =10$. Results for other $\tan\beta$ values are qualitatively similar. 
Then we scan over:
\bea
m_0:\ 0-15\ {\rm~TeV},\nonumber\\
m_{1/2}:\  0-2\ {\rm~TeV},\\
-2.5<\ A_0/m_0\ <2.5 .\nonumber
\label{eq:pspace}
\eea
We will show results for both $\mu >0$ and $\mu <0$.
For each solution generated, we require 
\begin{enumerate}
\item electroweak symmetry be radiatively broken (REWSB), 
\item the neutralino $\tz_1$ is the lightest MSSM particle, 
\item the light chargino mass obeys the LEP2 limit that
$m_{\tw_1}>103.5$~GeV~\cite{lep2},
\item $m_h=125\pm 2$~GeV (assuming $\pm 2$~GeV theory error in the $m_h$
calculation) in accord with the recent Higgs-like resonance discovery at
LHC~\cite{atlas_h,cms_h},
\item LHC search constraints on $m_{\tq}$ and $m_{\tg}$ are obeyed, where $m_{\tg}\agt 1$ TeV for $m_{\tg}\ll m_{\tq}$
and $m_{\tg}\agt 1.5$ TeV for $m_{\tg}\sim m_{\tq}$.\footnote{Explicit contours are shown in Ref. \cite{sugra}.}
\end{enumerate}

For mSUGRA, all GUT scale soft SUSY breaking scalar masses are equal to
$m_0$ while all gaugino masses equal $m_{1/2}$. In this case, from
Eq.~(\ref{eq:mHu}) we can calculate the $\delbg$ sensitivity co-efficients:
\bea 
c_{m_{1/2}} &=& (7.57 m_{1/2}-0.821 A_0 )(m_{1/2}/m_Z^2) ,\nonumber \\
c_{m_0} &=& 0.013 (m_0^2/m_Z^2),\nonumber \\ 
c_{A_0} &=& (0.44A_0-0.821 m_{1/2})(A_0/m_Z^2) ,\\ 
c_\mu &=& -2.18 (\mu^2/m_Z^2) \nonumber.  
\eea 
Notice that in this model, since $m_{H_u}=m_0(3)\equiv m_0$, there are large
cancellations in Eq.~(\ref{eq:mHu}) which suppress the contribution to
$c_{m_0}$. 

\begin{figure}[tbp]
\includegraphics[height=0.4\textheight]{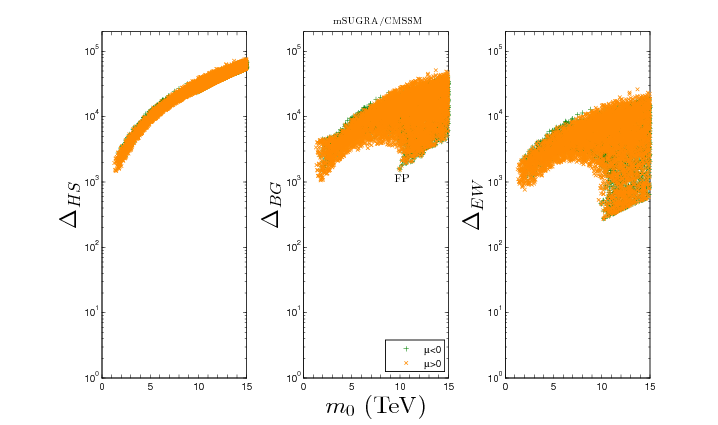}
\caption{Plot of $\Delta_{HS}$, $\Delta_{BG}$ and $\Delta_{EW}$ versus
$m_0$ from a scan over mSUGRA/CMSSM model parameter space for $\tan\beta =10$.
The location of the HB/FP regions is denoted FP in frame {\it b}).
\label{fig:sugm0}}
\end{figure}

In Fig. \ref{fig:sugm0}, we show $\Delta_{HS}$, $\Delta_{BG}$ and $\Delta_{EW}$ versus $m_0$ 
from our mSUGRA parameter space scan. In frame {\it a}), we see that $\Delta_{HS}$ is highly correlated with
$m_0$. This is to be expected since the larger $m_0$ becomes, the larger the top squark contributions are 
to $\delta m_{H_u}^2$. Thus, $\Delta_{HS}$ prefers the lowest $m_0$ values possible. We also see that
the minimal value of $\Delta_{HS}\sim 10^3$, corresponding to $\Delta^{-1}\sim 0.1 \%$ fine-tuning at best.
In frame {\it b}), we see that $\Delta_{BG}$ has a similar minimal value of $\Delta_{BG}\sim 10^3$, 
but the shape vs. $m_0$ is very different. 
One minimum occurs around $m_0\sim 2$ TeV while another minimum occurs at $m_0\sim 9$ TeV.
For $A_0\ne 0$, the contours of $\mu$ increase with $m_0$ and $c_\mu$ dominates $\Delta_{BG}$. 
At very high $m_0$,
one begins approaching what is known as the hyperbolic branch/focus point region\cite{ccn,fp} 
where $\mu$ decreases with increasing $m_0$: this causes the dip around $m_0\sim 9$ TeV and corresponds to
reduced fine-tuning even when scalar masses are very heavy\cite{fp}. 
Note that even though the min of $\Delta_{BG}$ drops around $m_0\sim 9$ TeV, 
the minimal value is still $\Delta_{BG}\sim 10^3$, or at best
$\sim 0.1\%$ EWFT. In frame {\it c}), we plot $\Delta_{EW}$ vs. $m_0$. For mSUGRA, $\mu^2\sim -m_{H_u}^2$ 
at the weak scale and since $\mu^2$ drops as one increases $m_0$ (for not too large $A_0$), then the HB/FP
region has the lowest $\Delta_{EW}$. Once $m_0$ exceeds $\sim 10$ TeV, then the $\Sigma_u^u$ terms dominate, 
and $\Delta_{EW}$ again increases with increasing top squark masses.
The min of $\Delta_{EW}$ is $\sim 250$, or $\sim 0.4\%$ fine-tuning in constructing $m_Z$. Thus, mSUGRA seems
rather highly electroweak fine-tuned under all three measures.

\begin{figure}[tbp]
\includegraphics[height=0.4\textheight]{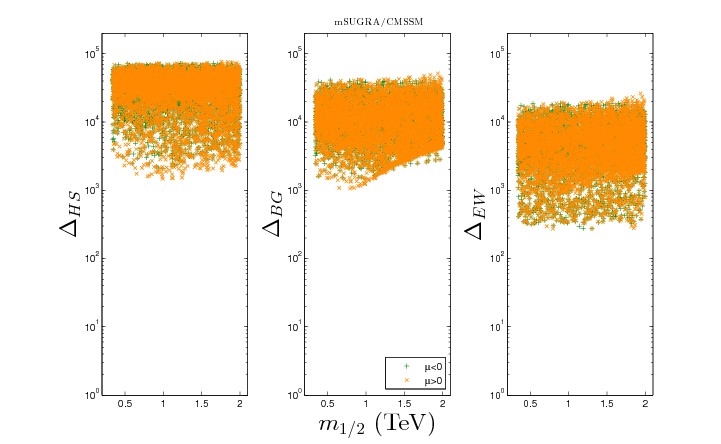}
\caption{Plot of $\Delta_{HS}$, $\Delta_{BG}$ and $\Delta_{EW}$ versus
$m_{1/2}$ from a scan over mSUGRA/CMSSM model parameter space for $\tan\beta =10$.
\label{fig:sugmhf}}
\end{figure}

In Fig. \ref{fig:sugmhf}, we show the three $\Delta$ measures vs. $m_{1/2}$. 
The min of $\Delta_{HS}$ is soft but occurs around $\sim 1$ TeV, as does the min of $\Delta_{BG}$. 
However, the distributions are really quite diffuse,
and for any $m_{1/2}$ value, a wide range of $\Delta_i$ values can occur. For $\Delta_{EW}$, there seems no
preference for any $m_{1/2}$ values, which is just a reflection that $\mu$ increases with 
$m_0$ and not $m_{1/2}$.

\begin{figure}[tbp]
\includegraphics[height=0.4\textheight]{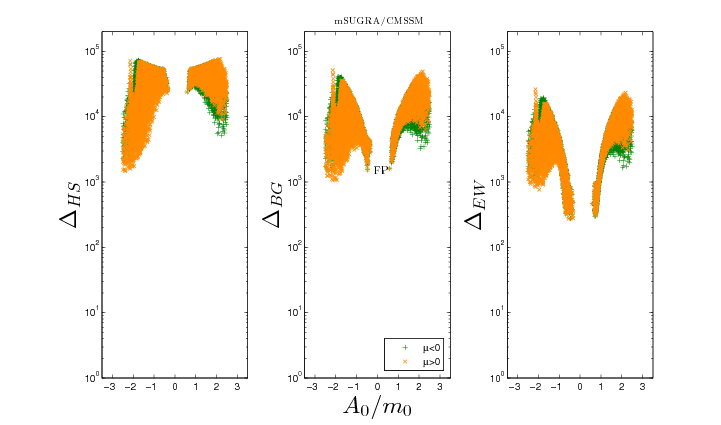}
\caption{Plot of $\Delta_{HS}$, $\Delta_{BG}$ and $\Delta_{EW}$ 
versus $A_0/m_0$ from a scan over mSUGRA/CMSSM model parameter
space for $\tan\beta =10$.
The location of the HB/FP regions is denoted FP in frame {\it b}).
\label{fig:A0}}
\end{figure}

In Fig. \ref{fig:A0}, we show the $\Delta$ measures vs. $A_0/m_0$. 
The first aspect of note is that no solutions occur 
for $A_0\sim 0$, which is because no solutions with $m_h\sim 123-127$ GeV can be found 
in the minimal stop mixing region.
The lowest $\Delta_{HS}$ values occur at largest $|A_0|$ values. This is because low $\Delta_{HS}$ prefers low
$m_0$, and low $m_0$ can only give $m_h\sim 123-127$ GeV for highly mixed stops. For $\Delta_{BG}$, 
one also gets a min at $A_0\sim 0.5m_0$. This is again the HB/FP region, where low $\Delta_{BG}$ is found
at high $m_0$, but at high $m_0$, not so much stop mixing is needed to obtain $m_h\sim 123-127$ GeV.
As in Fig. \ref{fig:sugm0}{\it c}), the min of $\Delta_{EW}$ is found for $|A_0|\sim 0.5m_0$, again in the
HB/FP region.

\begin{figure}[tbp]
\includegraphics[height=0.4\textheight]{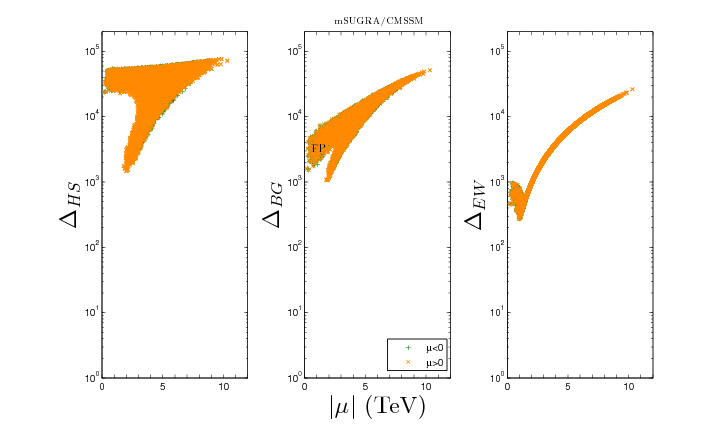}
\caption{Plot of $\Delta_{HS}$, $\Delta_{BG}$ and $\Delta_{EW}$ versus
$\mu$ from a scan over mSUGRA/CMSSM model parameter space for $\tan\beta =10$.
The location of the HB/FP regions is denoted FP in frame {\it b}).
\label{fig:sugmu}}
\end{figure}

In the mSUGRA model, the value of $m_{H_u}^2(m_{SUSY})$ is generated from its initial value $m_0$ at $m_{GUT}$
followed by RG evolution. The value of $\mu$ is then chosen by using Eq. \ref{eq:mZs} to determine
what $\mu^2(m_{SUSY})$ should have been in order to obtain the measured value of $m_Z$. 
In Fig. \ref{fig:sugmu}
we plot the $\Delta_i$ vs. $|\mu (m_{SUSY})|$. 
The lowest value of $\Delta_{HS}$ occurs around $\mu\sim 2$ TeV,
which correspond to values of $\mu$ where $m_0$ is minimal. The lowest values of $\mu$ in the 
$100-200$ GeV range come from the HB/FP region, but in this region $\Delta_{HS}$ is very large owing 
to the heavy top squarks. In frame {\it b}), we see $\Delta_{BG}$ is also split at low $\mu$, but this time
the higgsino region (HB/FP) with $\mu\sim 100-200$ GeV is only slightly more fine-tuned than the
lowest $\Delta_{BG}$ values. In frame {\it c}), the minimal $\Delta_{EW}$ occurs around $\mu\sim 1$ TeV, and
again the deep higgsino region (the region where the higgsino components of $\tz_1$ are dominant) 
has slightly larger values of $\Delta_{EW}$ owing to the large top squark masses
in the HB/FP region: these lead to large $\Sigma_u^u$. 
Thus, for the mSUGRA model, while all measures seem to favor low values of $\mu$, 
the lowest EWFT is not found in the deep higgsino (HB/FP) region, where a higgsino-like LSP is expected.

\begin{figure}[tbp]
\includegraphics[height=0.4\textheight]{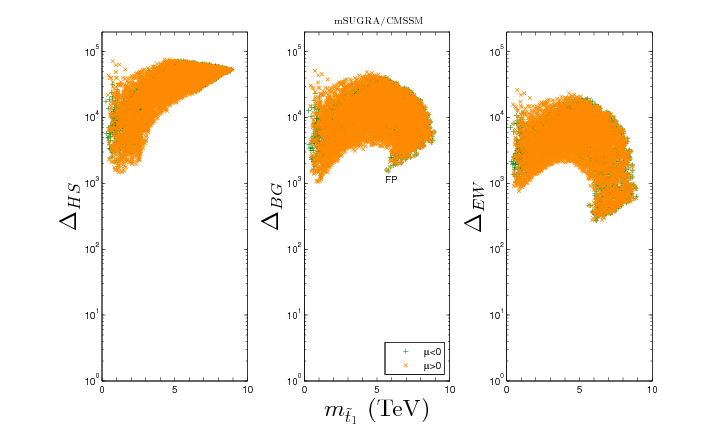}
\caption{Plot of $\Delta_{HS}$, $\Delta_{BG}$ and $\Delta_{EW}$ versus
$m_{\tst_1}$ from a scan over mSUGRA/CMSSM model parameter space for $\tan\beta =10$.
\label{fig:sug_mt1}}
\end{figure}

In Fig.~\ref{fig:sug_mt1} we show the $\Delta$ measures versus
$m_{\tst_1}$.  Here, we see that low $\Delta_{HS}$ does indeed occur
for relatively light top squarks, but with masses significantly above
the values advocated in Ref.~\cite{sundrum,ns}. Here, $m_{\tst_1}$ as low as
about $1$~TeV can be found; for lower $m_{\tst_1}$ values, very large
$|A_t|$ is required to satisfy the $m_h$ constraint by having very heavy
$\tst_2$; this, however, increases $\Delta_{HS}$. There are minima
of $\Delta_{BG}$ for $m_{\tst_1}\sim 1$~TeV and also in the HB/FP
region where $m_{\tst_1}\sim 6$~TeV. The measure $\Delta_{EW}$ also
shows two minima, with the lowest values being obtained in the HB/FP
region where stops are around 6~TeV.

\begin{center}
\begin{figure}[tbp]
\includegraphics[height=0.4\textheight]{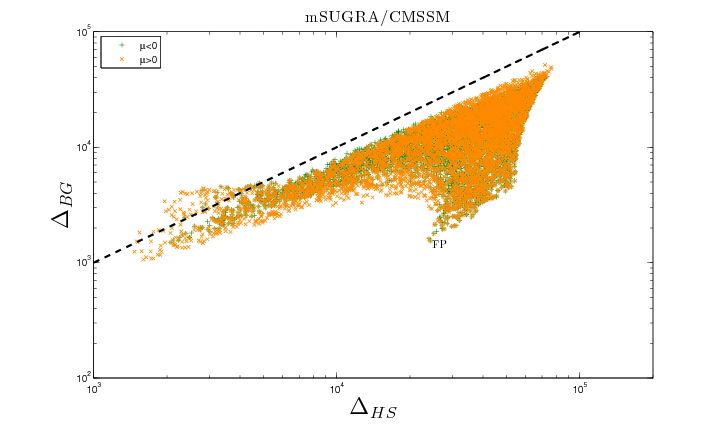}
\caption{Plot of $\Delta_{BG}$ versus $\Delta_{HS}$ from a scan over 
mSUGRA/CMSSM model parameter space for $\tan\beta =10$.
The dashed line denotes equal measures.
\label{fig:sug_bg_hs}}
\end{figure}
\end{center}

In Fig.~\ref{fig:sug_bg_hs}, for each parameter set generated, we plot
$\Delta_{HS}$ vs. $\Delta_{BG}$ to show the correlation between
the two measures. The dashed line shows where $\Delta_{BG}=\Delta_{HS}$.  
We see that the measures satisfy the inequality
in (\ref{eq:inequal2}),\footnote{The small number of points where
this inequality is violated is where $c_{m_{1/2}}$ determines $\delbg$;
in this case, the factor 2 arising from the fact we take the derivative
with respect to $m_{1/2}$ rather than $m_{1/2}^2$ plays an important
role. Indeed, it is easy to see that $\delhs\ge 2\delbg$ is always
satisfied.}  and further that the two measures are highly correlated: in
general, larger $\Delta_{HS}$ values also imply larger $\Delta_{BG}$. 
The exception occurs in the HB/FP region where, because of
correlations among the HS parameters, $\Delta_{BG}$ dips to very low
values even at large $\Delta_{HS}$.

\begin{figure}[tbp]
\includegraphics[height=0.4\textheight]{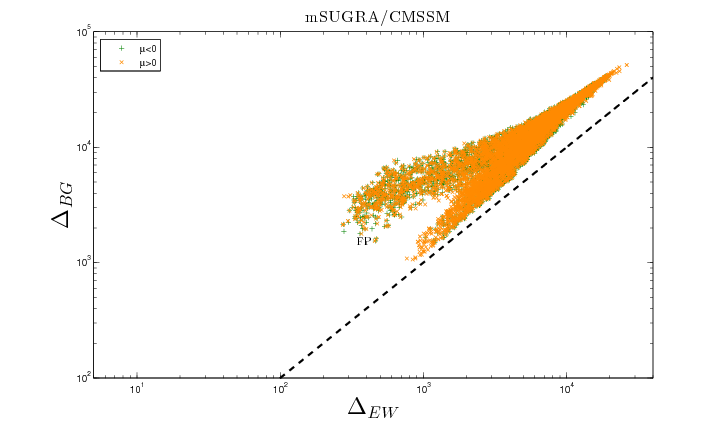}
\caption{Plot of $\Delta_{BG}$ versus $\Delta_{EW}$ from a scan over 
mSUGRA/CMSSM model parameter space for $\tan\beta =10$.
The dashed line denotes equal measures.
\label{fig:sug_bg_ew}}
\end{figure}

In Fig.~\ref{fig:sug_bg_ew}, we plot $\Delta_{EW}$ vs. $\Delta_{BG}$. 
We see that $\Delta_{EW} <\Delta_{BG}$, in accord with the
expectation in (\ref{eq:inequal}). Generally speaking, the
two measures are again well correlated; as before, 
the exception is  
the HB/FP region where $\Delta_{EW}$ decreases much more than $\Delta_{BG}$.

\section{The $\Delta_i$ in the NUHM2 model}
\label{sec:nuhm}

Next we turn to a scan over the two-extra-parameter non-universal Higgs
model NUHM2 defined by the parameter set,
\be m_0,\ m_{1/2},\ A_0,\ \tan\beta,\
\mu,\ m_A \ee 
and where again we fix $\tan\beta =10$. Then we scan over:
\bea
m_0:\ 0-15\ {\rm~TeV},\nonumber \\
m_{1/2}:\  0-2\ {\rm~TeV},\nonumber \\
-3.0 <\ A_0/m_0\ <3.0 ,\\
\mu :\ 0.1-1.5 \ {\rm~TeV},\nonumber \\
m_A:\  0.15-1.5\ {\rm~TeV} \nonumber.
\label{eq:nuhm2}
\eea
The points from this scan are shown by red pluses in the figures that
follow. We also performed a narrow scan with $\mu:\ 0.1-0.35$~TeV
denoted by blue crosses.  The constraints on the sparticle
masses are the same as in the mSUGRA
scan.

For $\Delta_{BG}$ in the NUHM2 model, the strong cancellation between
$m_{H_u}^2$ and the matter scalar mass terms in Eq.~(\ref{eq:mHu}) that was 
operative for mSUGRA no longer occurs.
Instead, the sensitivity coefficients are given by 
\bea 
c_{m_{1/2}} &=&
{\rm same\ as\ mSUGRA}, \nonumber\\ c_{m_0} &=& 1.336
(m_0^2/m_Z^2),\nonumber\\ c_{m_{H_u}} &=& -1.27
(m_{H_u}^2/m_Z^2),\nonumber\\ c_{m_{H_d}} &=& -0.053
(m_{H_d}^2/m_Z^2),\\ c_{A_0} &=& {\rm same\ as\ mSUGRA},\nonumber\\
c_\mu &=& {\rm same\ as\ mSUGRA}. \nonumber 
\eea 
The non-cancelling terms in NUHM2 now means that $\Delta_{BG}$ will
largely be driven by $c_{m_0}$ and $c_{m_{H_u}}$; an example is shown in Table \ref{tab:coeff}.
The sensitivity coefficient from $H_d$ is quite suppressed
compared to the $H_u$ term as expected for moderate-to-large $\tan\beta$.

\begin{figure}[tbp]
\includegraphics[height=0.4\textheight]{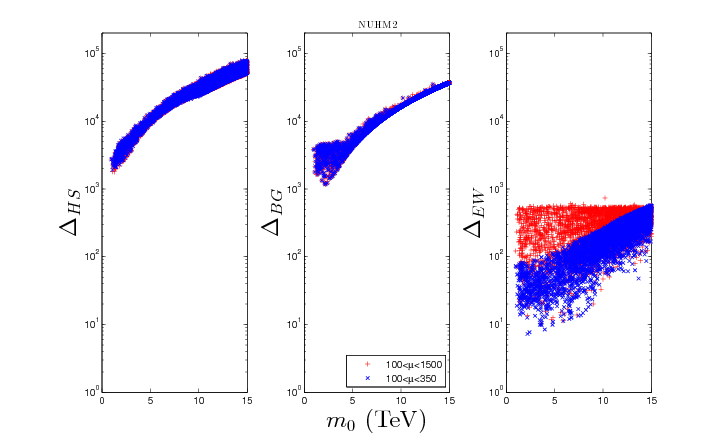}
\caption{Plot of $\Delta_{HS}$, $\Delta_{BG}$ and $\Delta_{EW}$ versus $m_0$ for the 
two  scans over NUHM2 model parameter space described in the text with $\tan\beta =10$.
The broad scan is denoted by red whilst the narrow scan with low $\mu$ is denoted by blue.
\label{fig:num0}}
\end{figure}

In Fig. \ref{fig:num0}, we plot the various $\Delta$s vs. $m_0$ for our scan of NUHM2 models.
From the plot, we see very different behaviors compared to Fig. \ref{fig:sugm0}. Both $\Delta_{HS}$
and $\Delta_{BG}$ are highly correlated with $m_0$, as may be expected in the $BG$ case if the
sensitivity coefficients are dominated by $c_{m_0}$. 
Minimal EWFT occurs at the lowest $m_0$  points available.
The minimal values of these two measures lie near $10^3$, similar to the mSUGRA case. 
The behavior of $\Delta_{EW}$ is very different. First, the minimal
value for $\Delta_{EW}$ from NUHM2 scan is around 7 ($\sim 14\%$ fine-tuning), 
which is about two orders of magnitude lower than the min from $\Delta_{HS}$ and $\Delta_{BG}$. 
These very low $\Delta_{EW}$ values occur in the radiatively-driven natural SUSY (RNS) 
scenario of Ref's. \cite{ltr,rns}.\footnote{In the radiatively-driven natural SUSY (RNS) model of 
Ref. \cite{ltr,rns}, $|\mu |$ is required $\sim 100-200$ GeV, $m_{H_u}^2$ is driven radiatively to
values $m_{H_u}^2\sim -m_Z^2$ and large mixing in the stop sector diminishes the radiative
corrections $\Sigma_u^u(\tst_{1,2})$ whilst lifting the value of $m_h$ to $\sim 125$ GeV.}
At tree-level, low $\Delta_{EW}$ is obtained for 1. low values of $\mu\sim m_Z$ and 2. low values of $m_{H_u}^2\sim m_Z^2$:
both these features can be realized, along with not-too-heavy stops, due to the extra parameter freedom enjoyed by NUHM2 models.
Furthermore, the distribution of $\Delta_{EW}$, while increasing with $m_0$, is only softly dependent on $m_0$, 
with minimal values of $\Delta_{EW}$ occurring in the $m_0\sim 2-5$ TeV range. 
This is because $m_0$ influences the top squark masses, which enter $\Delta_{EW}$ only at one-loop level.

\begin{figure}[tbp]
\includegraphics[height=0.4\textheight]{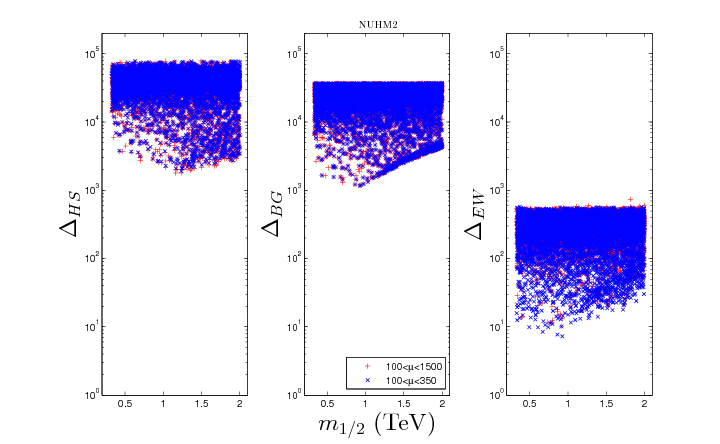}
\caption{Plot of $\Delta_{HS}$, $\Delta_{BG}$ and $\Delta_{EW}$ versus
$m_{1/2}$ from the two  scans over NUHM2 model parameter space described
  in the text with $\tan\beta =10$.
\label{fig:numhf}}
\end{figure}

In Fig.~\ref{fig:numhf}, we show the $\Delta_i$ vs. $m_{1/2}$. Here, as
in the mSUGRA case discussed before, all
 three  $\Delta_i$  exhibit only a weak dependence on
$m_{1/2}$, with the minimum of  $\Delta_i$ occuring around
$m_{1/2}\sim 1$~TeV. The results are independent of the range of $\mu$ 
that is scanned. 

\begin{figure}[tbp]
\includegraphics[height=0.4\textheight]{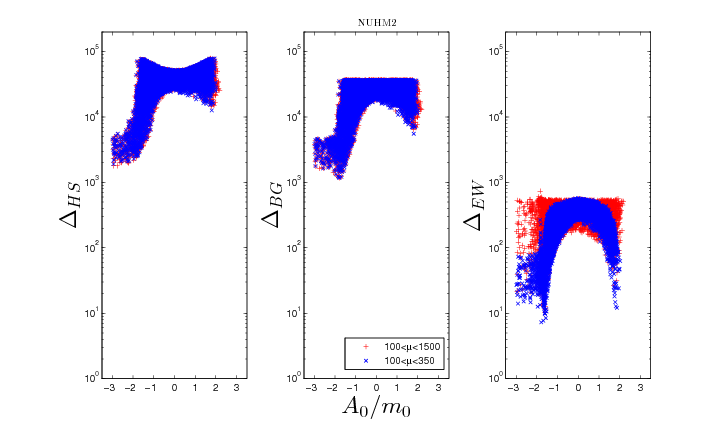}
\caption{Plot of $\Delta_{HS}$, $\Delta_{BG}$ and $\Delta_{EW}$ 
versus $A_0/m_0$ from the two scans over NUHM2 model parameter
space described in the text for $\tan\beta =10$.
\label{fig:nuA0}}
\end{figure}

In Fig.~\ref{fig:nuA0}, the values of $\Delta_i$ are shown versus
$A_0/m_0$.  Here, in contrast to the mSUGRA case,
we see no gap at small values of $A_0/m_0$ as
also noted in Ref.~\cite{h125}. The added freedom
to choose the Higgs mass parameters allows solutions with the observed
value of $m_h$. Both $\Delta_{HS}$ and
$\Delta_{BG}$ distributions exhibit  minima occurring at
large $|A_0|$ values. The lowest value for $\Delta_{HS,BG}$ occurs at
$A_0\sim -(2-3)m_0$.  For this sign of $A_0$, it is much easier to
generate $m_h\sim 125$~GeV at low $m_0$ values where the $\Delta_{HS,BG}$ are lowest. 
The distribution in $\Delta_{EW}$ also has
minima at large $A_0$, although the minima tend to occur around $A_0\sim
\pm 1.6 m_0$.

\begin{figure}[tbp]
\includegraphics[height=0.4\textheight]{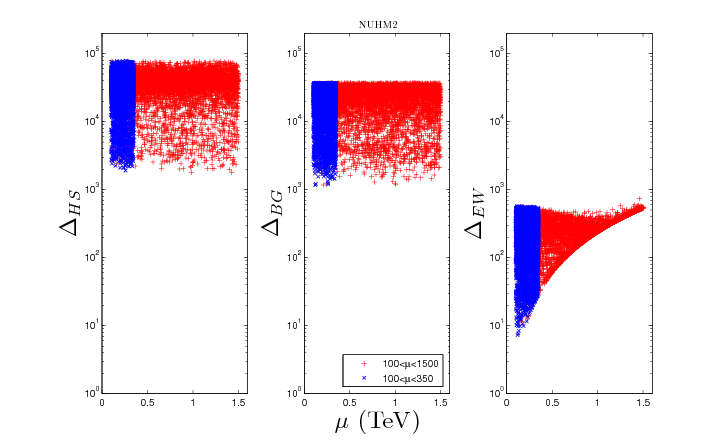}
\caption{Plot of $\Delta_{HS}$, $\Delta_{BG}$ and $\Delta_{EW}$ versus
$\mu$ from the two  scans over NUHM2 model parameter space described in
  the text
for $\tan\beta =10$.
\label{fig:numu}}
\end{figure}

In Fig.~\ref{fig:numu}, we show the three $\Delta_i$ measures vs. $\mu$.
In this case, we see that neither $\Delta_{HS}$ nor $\Delta_{BG}$ have any preferred $\mu$ value. 
This is because $\mu$ does not
enter the evolution of $m_{H_u}^2$ (or any other soft SUSY breaking
parameter) in the case of $\Delta_{HS}$, and $c_\mu$ is never the
maximal sensitivity coefficient in $\Delta_{BG}$.  The situation is
quite different for $\Delta_{EW}$. In this case, we see a tight
correlation with the low $\Delta_{EW}$ values preferring the lowest
values of $\mu$ that are phenomenologically allowed, {\it i.e.} those
closest to $m_Z$. In this case, low $\Delta_{EW}$ has a strong
preference for a set of light higgsinos $\tw_1^\pm, \tz_{1,2}$ of which
the $\tz_1$ would be a higgsino-like WIMP candidate. Note, however, that
the gaugino components of $\tz_1$ cannot get too small since then large
gaugino masses would increase $m_{\tst_{1,2}}$, thus increasing the
radiative corrections $\Sigma_u^u$. We thus conclude that $\tz_1$ has
substantial higgsino {\it and} gaugino components, giving it an
observable spin-independent direct detection cross section $\sigma^{SI}
(\tz_1 p)$ at ton-sized detectors \cite{bbm}.  Also, the various
higgsinos would likely be visible at a linear $e^+e^-$ collider
operating with $\sqrt{s}\sim 0.25-1$~TeV, although these would be
difficult to directly observe at the LHC if gluinos are heavier than
1.5-2~TeV \cite{bbh} because then, the $\tw_1/\tz_2-\tz_1$ mass gap
becomes too small. For the low $|\mu|$ models, the $pp\to\tw_2\tz_4\to W^\pm W^\pm +\eslt$ signal
may be observable at the (high luminosity) LHC if the heavy wino-like
$\tw_2$ and $\tz_4$ have masses up to about 800~GeV \cite{lhcltr}.

\begin{figure}[tbp]
\includegraphics[height=0.4\textheight]{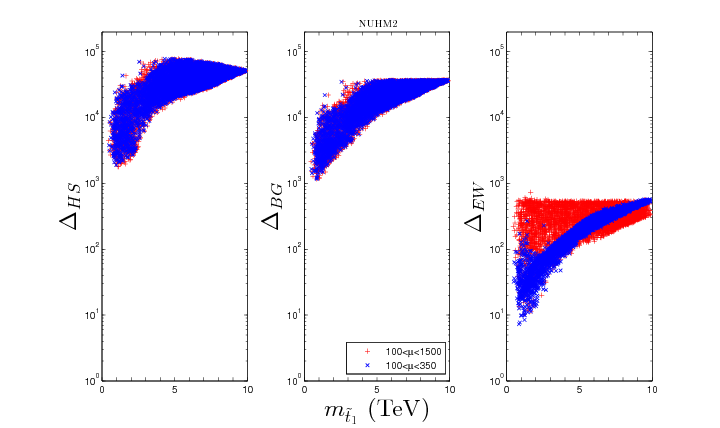}
\caption{Plot of $\Delta_{HS}$, $\Delta_{BG}$ and $\Delta_{EW}$ versus
$m_{\tst_1}$ for the two  scans over NUHM2 model parameter space
  described in the text for $\tan\beta =10$.
\label{fig:nu_mt1}}
\end{figure}

In Fig.~\ref{fig:nu_mt1} we show the distribution of $\Delta_i$ values
vs. $m_{\tst_1}$.  Here, we find all three measures concur that the
lowest fine-tuning is found for the lowest values of $m_{\tst_1}$ which lead to
$m_h\sim 125$~GeV. These tend to lie in the vicinity of $m_{\tst_1}\sim
1$~TeV, well beyond current bounds from LHC. Although 
we have not shown it here, we have checked that the corresponding value of
$m_{\tst_2} \sim 2$~TeV. 

\begin{figure}[tbp]
\includegraphics[height=0.4\textheight]{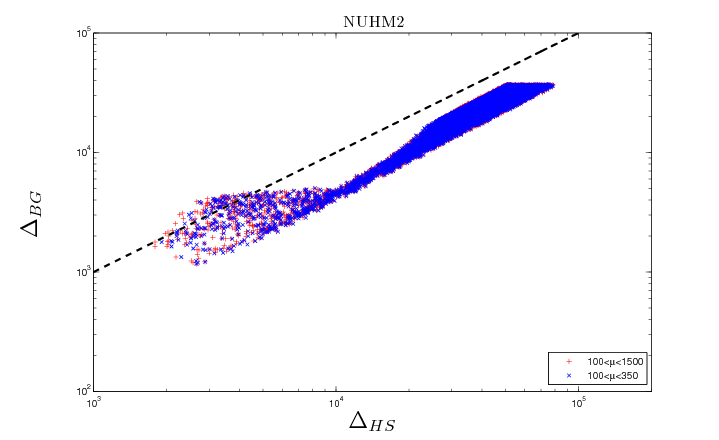}
\caption{Plot of $\Delta_{BG}$ versus $\Delta_{HS}$ for the two scans over 
NUHM2 model parameter space described in the text for $\tan\beta =10$.
\label{fig:nu_bg_hs}}
\end{figure}

Figure \ref{fig:nu_bg_hs} shows the correlation between $\Delta_{HS}$ and $\Delta_{BG}$ 
from the scan over NUHM2 parameter space. For
NUHM2, these two measures are highly correlated, and once again we see
that the inequality (\ref{eq:inequal2}) is broadly satisfied, and
further that all points satisfy $\delbg \le 2\delhs$ as mentioned
earlier.

\begin{figure}[tbp]
\includegraphics[height=0.4\textheight]{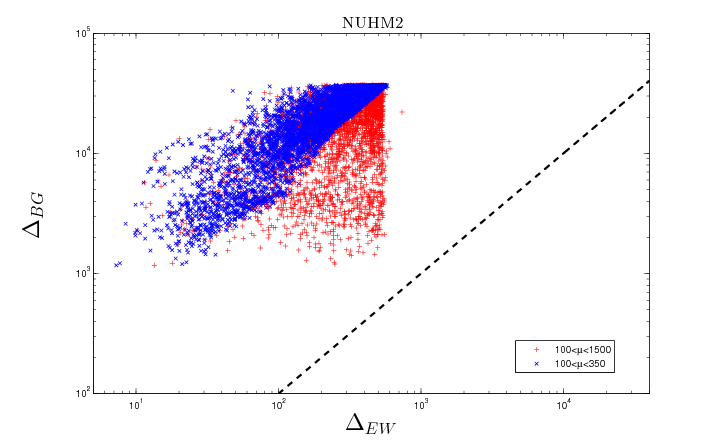}
\caption{Plot of $\Delta_{BG}$ versus $\Delta_{EW}$ for the two
scans over NUHM2 model parameter space described in the text for $\tan\beta =10$.
\label{fig:nu_bg_ew}}
\end{figure}

Finally, in Fig.~\ref{fig:nu_bg_ew}, we plot $\Delta_{BG}$
vs. $\Delta_{EW}$.  Once again, we see that (\ref{eq:inequal}) is
satisfied. Whereas these two $\Delta$s were highly correlated in the
mSUGRA case (except for the points in the HB/FP region), for NUHM2 they
are much less so in that points with lowest $\Delta_{BG}$ may have
$\Delta_{EW}$ ranging from its minimum at $\sim 7$ all the way up to
near its maximum. This is because points with very large $\mu$ values
can have very low values of $\delbg$ because, as we have already noted,
$c_\mu$ is never maximal in the various sensitivity coefficients. 

A comparison of the three $\Delta$s for each of three models--  mSUGRA, NUHM2 and pMSSM\footnote{
The pMSSM, or phenomenological MSSM, is the MSSM defined with 19 free weak scale parameters.}--
is shown in Table \ref{tab:Deltas}. 
For mSUGRA, we take $m_0=9993.4$ GeV, $m_{1/2}=691.7$ GeV, $A_0=-4788.6$ GeV and
$\tan\beta =10$. The mSUGRA output values of $\mu =309.7$ GeV and 
$m_A=9859.9$ GeV serve as NUHM2 inputs.
The weak scale outputs of mSUGRA and NUHM2 serve as pMSSM inputs so that
all three models have exactly the same weak scale spectra. 
From the Table, we see under $\Delta_{HS}$, the mSUGRA and NUHM2 models are both highly
fine-tuned since $\Delta_{HS}$ mainly depends on the change $\delta m_{H_u}^2$ in running from
$m_{GUT}$ to $m_{SUSY}$. For the pMSSM, since $\Lambda =m_{SUSY}$, then $\Delta_{HS}$ collapses
to $\Delta_{EW}$. 
For the measure $\Delta_{BG}$, we obtain maximal EWFT within the NUHM2 model since here we have
a large set of uncorrelated parameters at the scale $\Lambda =m_{GUT}$. In mSUGRA, with fewer
parameters due to $m_{H_u}=m_{H_d}\equiv m_0$, the additional correlations allow for a
collapse in EWFT by an order of magnitude. If additional parameter correlations are present, 
{\it e.g.} relating $m_0$ with $A_0$ and $m_{1/2}$, then it is possible that $\Delta_{BG}$
collapses even further to near its lower limit given by $\Delta_{EW}$. 
Under $\Delta_{EW}$, which is model-independent (within the MSSM), then all three models
have identical values of EWFT: $\Delta_{EW}=462$.
\begin{table}
\begin{center}
\begin{tabular}{|l|r|r|r|}
\hline
model & $\Delta_{HS}$  &  $\Delta_{BG}$ & $\Delta_{EW}$ \\
\hline
\hline
mSUGRA    & 24302 &  1540  & 462 \\
NUHM2     & 24302 &  16041 & 462 \\
pMSSM     & 462   &  462   & 462 \\
\hline
\end{tabular}
\caption{Values of $\Delta_{HS}$, $\Delta_{BG}$ and $\Delta_{EW}$ 
for the mSUGRA/CMSSM, NUHM2 and pMSSM models. 
For mSUGRA, we take $m_0=9993.4$ GeV, $m_{1/2}=691.7$ GeV, $A_0=-4788.6$ GeV and
$\tan\beta =10$. The mSUGRA output values of $\mu =309.7$ GeV and 
$m_A=9859.9$ GeV serve as NUHM2 inputs.
The weak scale outputs of mSUGRA and NUHM2 serve as pMSSM inputs so that
all three models have exactly the same weak scale spectra. 
\label{tab:Deltas}}
\end{center}
\end{table}

\section{Interpretation in terms of an ultimate theory} 
\label{sec:meta}

In this paper, we have computed three measures of electroweak naturalness and applied them
to two popular models: mSUGRA and NUHM2.
We have argued that $\Delta_{HS}$ produces an overestimate of EWFT due to a separation
of {\it dependent} terms $m_{H_u}^2 (\Lambda )$ and $\delta m_{H_u}^2$. These terms contain 
large correlated cancellations  since the larger $m_{H_u}^2 (\Lambda )$ becomes, the larger
is the radiative correction $\delta m_{H_u}^2$. In fact, the large negative correction
contained in $\delta m_{H_u}^2$ is exactly what is required to cause a radiatively generated
breakdown in electroweak symmetry. 
The measure $\Delta_{BG}$ avoids this problem by evaluating the combination 
$m_{H_u}^2(\Lambda )+\delta m_{H_u}^2 =m_{H_u}^2 (m_{SUSY} )$ in terms of fundamental model parameters.
By invoking HS models with increasingly constrained parameter sets, the EWFT in $\Delta_{BG}$
can be seen to collapse. An explicit demonstration occurs in moving from the six-parameter 
NUHM2 model to the four-parameter mSUGRA model in the HB/FP region: in this case, 
much lower values of $\Delta_{BG}$ are generated in the region of heavy stop masses than might 
otherwise be expected under $\Delta_{HS}$.

At this point, we should note that few authors would be willing to consider either mSUGRA or NUHM2
as fundamental theories. Instead, they are to be viewed as effective field theories whose
range of validity may extend up to $\Lambda =m_{GUT}$.
An often unstated assumption is that most authors hypothesize the existence of an 
overarching ultimate theory-- perhaps string theory-- 
whose low energy limit for $Q<\Lambda =m_{GUT}$ is the MSSM but wherein the 
high scale soft terms are all correlated (here referred to as the UTH). 
Such an UTH would have fewer free parameters, and perhaps even no parameters at all: 
in the latter case, the soft terms might all be determined in terms of the fundamental Planck scale $M_P$. 
The UTH might be contained within the more general effective SUSY theories popular in the literature.
In the case of $\Delta_{BG}$, the measure of EWFT can be overestimated by evaluating $\Delta_{BG}$ 
within the effective theories instead of within the UTH. 
Indeed, it is not even clear if $\Delta_{BG}$ has any meaning for an UTH with no free parameters. 

\begin{figure}[tbp]
\includegraphics[height=0.4\textheight]{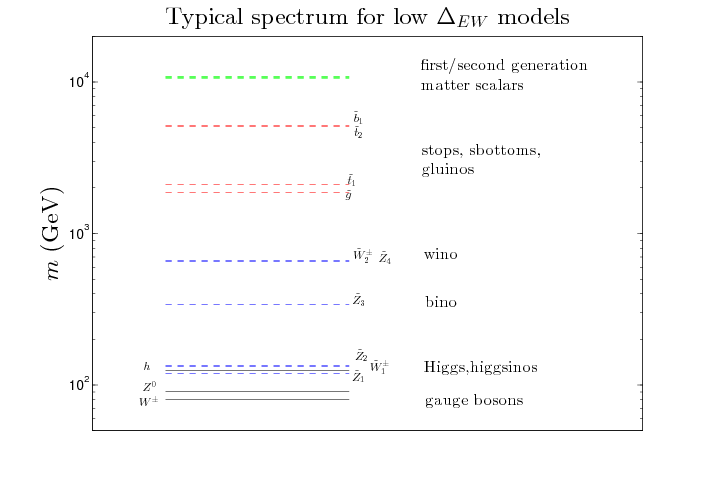}
\caption{Typical sparticle mass spectrum from SUSY models with low $\Delta_{EW}$.
Such a spectrum might be expected to result from an UTH with low true EWFT.
\label{fig:mass}}
\end{figure}

The measure $\delew$ allows for the possibility of parameter correlations which should be present 
in the UTH and, since it is model-independent, leads to the same value of $\delew$ for the
effective theories as should occur for the UTH.
In the course of this work, we have found that the well-known mSUGRA/CMSSM model is 
fine-tuned under all three measures. As such, it is unlikely to contain the UTH. 
The non-universal Higgs model NUHM2 appears fine-tuned with $\Delta_{HS,BG} \gtrsim 10^3$. 
But since $\delew$ can be as small as 7 (corresponding to 14\% fine-tuning or one part in 10), 
it may contain the UTH for selected parameter choices which allow for low $\Delta_{EW}$.
In other words, a model with derived parameters leading to low $\Delta_{EW}$ would 
also have low true EWFT.
In the case of NUHM2, which preserves the elegant SUSY and GUT features, the UTH should lead to
typical mass spectra shown in Fig. \ref{fig:mass}. For even more general frameworks ({\it e.g.}
those with non-universal gaugino masses) other spectra with low $\Delta_{EW}$ 
are also possible\cite{shafi,sug19}.

\begin{figure}[tbp]
\includegraphics[height=0.4\textheight]{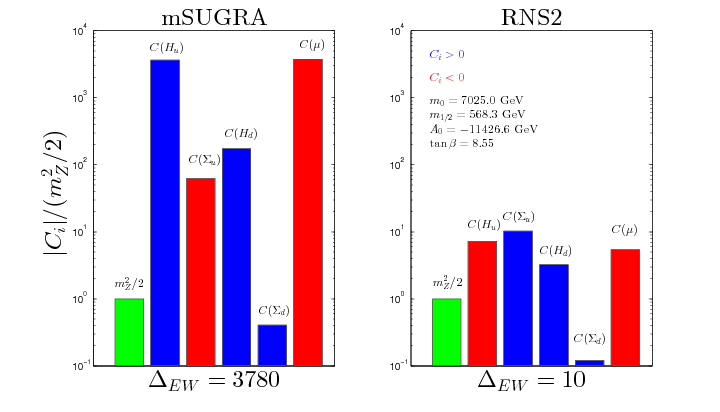}
\caption{Plot of contributions to $m_Z^2/2$ from the  mSUGRA/CMSSM model with 
parameters as listed, and also for RNS2 benchmark point with the same $m_0$, $m_{1/2}$, 
$A_0$ and $\tan\beta$ values, but with $\mu=150$ GeV.
Red bars denote negative contributions while blue bars denote positive contributions.
\label{fig:bar}}
\end{figure}

The measure $\Delta_{EW}$ is model independent in the sense that any high scale model giving rise
to look-alike spectra at the weak scale will have the same value of $\Delta_{EW}$. It is also
most intimately connected with data, in that it requires natural generation of $m_Z=91.2$ GeV
while maintaining LHC Higgs mass and sparticle mass constraints. In this sense, models with low
$\Delta_{EW}$ solve what is known as the Little Hierarchy Problem: how can it be that
$m_Z$ and $m_h\sim 100$ GeV while sparticle masses are beyond the TeV scale. The answer is that the
SUSY models must have low Higgsino mass $\mu\sim 100-200$ GeV, they must generate
$m_{H_u}^2\sim -(100-200)$ GeV at the weak scale (always possible in NUHM models) and there
must be large mixing in the top-squark sector with TeV-scale top squarks. 
An example may be seen in Fig. \ref{fig:bar}. Here we show the various scalar potential contributions
to $m_Z$ scaled to $m_Z^2/2$ for $m_0=7025$ GeV, $m_{1/2}=568.3$ GeV, $A_0=-11426.6$ GeV and $\tan\beta =8.55$
(benchmark point RNS2 from Ref. \cite{ltr}). 
Red bars denote negative contributions while blue bars denote positive contributions.
In frame {\it a}), the situation is shown for the mSUGRA model 
(parameters as above with $m_{H_u}=m_{H_d}=m_0$) where
$m_{H_u}^2$ is driven to large negative values at the weak scale. 
The value of $\mu^2$ must be dialed in (fine-tuned) so that a large, 
unnatural cancellation between $m_{H_u}^2$  and $\mu^2$ is needed to gain a $Z$ mass of just 91.2 GeV.
In frame {\it b}), we show the case for radiatively-driven natural SUSY with the same 
parameters as mSUGRA but with $\mu =150$ GeV and where now $m_{H_u}(\Lambda )\ne m_{H_d}(\Lambda )\ne m_0$.
All contributions are now roughly comparable to $m_Z^2/2$ so that in this case it is easy to understand why 
$m_Z$ and $m_h$ both naturally occur around $\sim 100$ GeV. 

\begin{figure}[tbp]
\includegraphics[height=0.4\textheight]{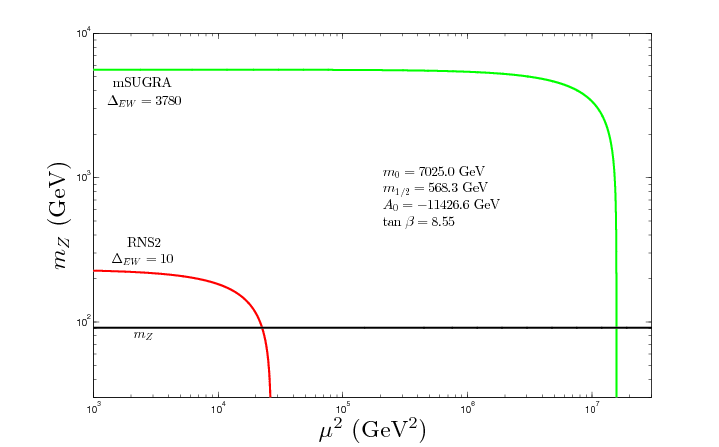}
\caption{Plot of $m_Z$ vs. $\mu^2$ for mSUGRA/CMSSM with 
parameters as listed. We also show $m_Z$ vs. $\mu^2$ in the NUHM2 model.
\label{fig:mzmu}}
\end{figure}

This can be further illustrated in Fig. \ref{fig:mzmu}, where we adopt all weak scale parameters from the two benchmark
models except $\mu^2$, but then plot the value of $m_Z$ as is generated by varying $\mu^2$. We see in the mSUGRA
case that one would naturally expect $m_Z\sim 6$ TeV instead of 91.2 GeV. One must finely tune $\mu^2$
to very high precision to generate $m_Z=91.2$ GeV. In the RNS2 case, $m_Z$ is expected to lie around 
$200$ GeV, and it is not so far fetched that it turns out to be 91.2 GeV, which still requires $\sim 10\%$
fine-tuning of $\mu^2$. 

\section{Conclusions}
\label{sec:concl}

Our conclusions are summarized as follows.
\begin{itemize}
\item The measure $\Delta_{HS}$, which essentially measures $\delta m_{H_u}^2/(m_h^2/2)$ or
alternatively $\delta m_{H_u}^2/(m_Z^2/2)$, overestimates EWFT by omitting the non-independent
value of $m_{H_u}^2(\Lambda )$. This can be rectified by instead using the combined term 
$(m_{H_u}^2(\Lambda )+\delta m_{H_u}^2)/(m_Z^2/2)$ as occurs in $\Delta_{BG}$ and $\Delta_{EW}$.
\item $\Delta_{BG}$ measures fractional change in $m_Z^2$ against fractional change in model parameters.
As such, it is by definition model-dependent. To interpret $\Delta_{BG}$, 
we introduce the concept of an over-arching UTH with few or even no free parameters.
By applying $\Delta_{BG}$ to the more general effective theories which contain the UTH, then
large cancellations due to correlated high scale parameters are missed, leading to an overestimate in EWFT.
An example is shown where mSUGRA functions as a four-parameter UTH contained within the six-parameter NUHM2.
The correlated parameters $m_{H_u}^2=m_{H_d}^2\equiv m_0^2$ lead to large cancellations in the scalar sector in the
well-known HB/FP region. 
\item The model-independent measure $\Delta_{EW}$ is obtained as the limit as $\Lambda\to m_{SUSY}$ of
$\Delta_{HS}$ and $\Delta_{BG}$. It measures how likely the weak scale $\mu$ parameter, 
soft terms and radiative corrections can conspire to yield $m_Z,\ m_h\sim 100$ GeV 
without large uncorrelated cancellations (fine-tuning). 
(Typically, in models with large $\delew$, a value of $\mu^2$ must be dialed in (fine-tuned) 
so that a large,  unnatural cancellation between $m_{H_u}^2$  and $\mu^2$ is 
required to obtain a $Z$ mass of just 91.2 GeV.)

Since it is model independent and depends only on the weak scale spectra which is generated, 
it will produce the same value for the UTH as it would for various effective theories which contain the UTH.
While mSUGRA is fine-tuned under all three measures, implying it is unlikely to contain the UTH, 
values of $\Delta_{EW}$ below 10 can be found for the NUHM2 model, indicating fine-tuning to one part in 10.
The weak scale spectra from NUHM2 which yield $\Delta_{EW}\sim 10$ would be a good candidate for what may be 
expected from an UTH including SUSY/GUT relations with low true EWFT. Such models are characterized by 
light higgsinos $m_{\tw_1},\ m_{\tz_{1,2}}\sim 100-300$ GeV which can elude searches at LHC14\cite{wp2}, 
but which could easily be discovered at an $e^+e^-$ collider with $\sqrt{s}\sim 500-600$ GeV\cite{Berggren:2013vfa}.
\end{itemize}

Our overall lesson is that the conventional measures $\Delta_{HS}$ and $\Delta_{BG}$ tend to overestimate-- 
often by orders of magnitude-- the EWFT needed for supersymmetry theory. 
In contrast, as discussed in Ref. \cite{rns}, the measure $\Delta_{EW}$ has the properties of being 
model-independent, conservative, measureable, unambiguous, predictive, falsifiable and
simple to calculate. In virtue of these qualities-- and in light of our current lack of knowledge of the UTH--
the quantity $\Delta_{EW}$ appears to be the correct measure of EWFT to apply to the effective theories which
might contain the UTH.
In models such as NUHM2 which allow for $\Delta_{EW}$ as low as $\sim 10$, 
then\cite{altarelli} 
``the SUSY (GUT) picture $\cdots$ remains the standard way beyond the Standard Model.''
Target spectra for model builders intent on constructing the UTH are provided in Fig.~\ref{fig:mass}.

\section*{Acknowledgments}

We thank A. Mustafayev and X. Tata for discussions.
This work was supported in part by the US Department of Energy, Office of High
Energy Physics.

%

%
\end{document}